# Computing with bricks and mortar:
# Classification of waveforms with a doped concrete blocks

Dawid Przyczyna[ab]*, Maciej Suchecki[ab], Andrew Adamatzky[c], Konrad Szaciłowski[a]*

[a]AGH University of Science and Technology, Academic Centre for Materials and Nanotechnology, al. Mickiewicza 30, 30-059 Kraków, Poland

[b]AGH University of Science and Technology, Faculty of Physics and Applied Computer Science, al. Mickiewicza 30, 30-059 Kraków, Poland

[c]University of the West of England, Unconventional Computing Lab, Department of Computer Science and Creative Technologies, Bristol BS16 1QY, United Kingdom

*Corresponding authors: dawidp@agh.edu.pl, szacilow@agh.edu.pl

**Abstract**

We present results showing the capability of concrete-based information processing substrate in the signal classification task in accordance with *in materio* computing paradigm. As the Reservoir Computing is a suitable model for describing embedded *in materio* computation, we propose that this type of presented basic construction unit can be used as a source for *"reservoir of states"* necessary for simple tuning of the readout layer. In that perspective, buildings constructed from computing concrete could function as a highly parallel information processor for smart architecture. We present an electrical characterization of the set of samples with different additive concentrations followed by a dynamical analysis of selected specimens showing fingerprints of memfractive properties. Moreover, on the basis of obtained parameters, classification of the signal waveform shapes can be performed in scenarios explicitly tuned for a given device terminal.

1. **Introduction**

In developed countries, technology begins to encompass more and more aspects of life. Approximately 87% of humanity has access to electricity, according to the International Energy Agency [1]. In turn, less than half of the population has a continuous access to the Internet [2]. Both these percentages increase every year, indicating progressing technological advancement of the human race. Nowadays, a technology that surrounds people with devices connected to the internet – so-called Internet-of-things (IoT) – is beginning to gain increasing recognition [3]. It can take the form of "smart home", with connected home appliances, heating, lighting and wearables of inhabitants with a smartphone or smart speakers in an attempt to increase comfort and security of human life (e.g. in the form of "elder care"). Other applications of IoT include healthcare, transportation, manufacturing, agriculture, or military. The global market for IoT was valued at 164mld $ in 2018 [4] and it is predicted that the global market of "smart homes" can reach 58mld $ in 2020 [5]. At the same time broadband access to the Internet as well as processing and storage of huge amount of data is extremely important. Fast information processing and storage, however, is an extremely energy-demanding technology. Therefore at least some of the data processing should be delegated into substrates other than silicon, operatingmuch slower, but at the same time consuming less energy. Alternatively, the waste heat produced during computing can be utilized for heating purposes in colder seasons. This may help to reduce the carbon footprint of computing, which nowadays accounts for 3.2% of the total anthropogenic carbon dioxide emissions [6].

Combination of ideas of *in materio* computing [7-11] and smart houses [12-15]

immediately leads to the concept of computational concrete – smart material combining construction and information processing features. If successful, such material should render each building an energy-efficient supercomputing device. What if walls would not only support the roof, but at the same time perform advanced, decentralized and distributed computation? Each building block would sense itself and the environment, monitor safety of the construction, environmental pollution, and interact with humans in an intelligent way. This far-fetched vision has been already proposed and supported with some preliminary experimental and theoretical investigations [16]. Selection of concrete as a computational medium seems shocking at the first glance. On the other hand various unorthodox substrates have been already reported to perform advanced computation, including liquid marbles [17, 18], slime molds [19, 20], mycelia and fungi [21, 22], algae [23] and photochromic solutions [24, 25]. In principle, any physical system of sufficiently complex, structure, dynamics, and responsiveness to external stimuli can be utilized for information processing [26]. In view of the above, the choice of concrete as a ubiquitous computational medium seems reasonable. Furthermore, concrete is easily and readily prepared and fabricated in all sorts of shapes and structural systems. Its great simplicity lies in the fact that its constituents are ubiquitous and are readily available almost anywhere in the world. As a result of its ubiquity, functionality and flexibility, it has become by far the most popular and widely used construction material in the world. It is particularly suitable for nano- or micro-modifications due to its peculiar internal structure. The ingredients can be selected, proportioned and engineered to produce a concrete of a specific strength and durability or other multifunctional properties, so it is 'fit for purpose' for the job for which it is intended [27]. It can be produced in the form of precast products or as ready-mixed concrete, which is delivered in the familiar rotating concrete lorry. Currently, ingredients are optimised to make concrete strong, light-weight, low-thermally conductive, and durable when exposed to the environment. However new investigations are focused on concrete with embedded sensing [28-32].

Current IoT technology includes a broad set of topics such as sensors, embedded systems and machine learning (ML). ML methods can be used to improve the functioning of intelligent infrastructure through the prediction of action of inhabitants based on their day cycle or increase the security of the whole system. [33, 34] This is done from the software side, treating the building structure only as a skeleton to ensure its durability and thermal insulation. Through the use of efficient ML methods, such as Reservoir Computing (RC), it becomes possible to develop intelligent infrastructure based on the building blocks capable of embedded, distributed information processing [16].

RC paradigm can be regarded as an extension of artificial neural networks (ANN) encompassing in its framework various physical substrates and processes [35-37]. Its main strength is the so-called "reservoir of states" possessing rich configuration state space of internal dynamics and performing nonlinear transformation of input signals. Thanks to its operation, simplification of the training process of ANN can be achieved, as probing of a reservoir at the readout layer is the only part of the system that needs tuning [38, 39]. Probing different features of the reservoir can enable the implementation of pattern recognition, assuming that the given configuration state space is diverse enough [40].

It has been shown by Wlaźlak *et.al.* that a pure hardware RC system based on single memristive nonlinear node operating in the delayed feedback loop can be used in the simple classification of signal amplitudes [41, 42]. More complex RC setup based on memristor array (supporting reservoir of states) with ANN software readout was studied as an image recognition system [43] and similar systems were considered for waveform recognition [44]. Therefore appropriate doping, which can induce memristive properties in concrete-based materials is desired. In our recent work, we have suggested the possibility of implementing RC concepts based on hybrid construction material – a "computing concrete" based infrastructure, that could potentially work as a massively interconnected parallel processor [16]. This assumption was drawn on the basis of rich and nonlinear responses of the device to the electrical stimulation. It was theorized that highly nonlinear response arose due to many different pathways for charge carriers and superimposition of capacitive behavior of the device with internal ionic movement. Buildings based on this type of embedded hardware could then possess multisensory properties and support forms of information processing.

Herein we present a hardware structure based on a dual system of computing/regular concrete samples. Impedance spectroscopy and cyclic voltammetry were employed to firstly characterize conductivity and electrical response of the set of samples. Further analysis was performed for the sample where memristive traces were found in CV measurements. The signals of different shapes and several frequencies were applied to two terminals of the device, were subjected to mixing in the sample, and then collected from the third terminal. The system responses were then characterized using a set of dynamical parameters describing its complexity, chaotic nature and several fractal dimensions of registered time series. The parameters obtained for the dual system show a rich configuration space, which makes the system a suitable platform for Reservoir Computing. Analysis and processing of obtained information allow for classification of signal shapes – between sinusoidal, triangle and square. It can be accomplished through several variants in the decision tree manner or depending on specific characteristics of a given readout. Three different sets of classification criteria has been established, all these sets yield indentical waveform classification performance on the basis of various statistical indices and dynamic parameters of recorded time series.

2. **Experimental**

The base material used in this experiment was ready-to use concrete mix procured from Leroy Merlin and steel shavings supplied by POCh (Poland). Antimony sulfoiodide nanowires (SbSI) were synthesized in the following procedure. The reactants weighed and added in a ratio of 1g Sb, 0.265g S and 1g I. All reactants were mixed in a 100ml flask using 50ml isopropanol as a solvent. The whole was placed in the ultrasonic bath previously heated to 50° C for 6 hours. The resulting product was purified by five-fold centrifugation for 5 minutes at 5000 rpm and washed three times with isopropanol and 2-fold with water after that product was left to dry.

The reference sample consisting of only concrete, as well as modified samples additionally containing 1%, 5% and 10% of either SbSI, steel shavings, or half and half mixture by weight of both, were created using the following steps. In the bottom of a plastic container, holes 1cm apart were made, creating a 3×3 grid. Those holes served as an insertion point for silver wires that would go through the bulk of the material. After preparing the mold, the material was poured in. The whole was firmly shaken to remove pockets of air and allow content to settle within the container. Water was poured until all concreate was sufficiently saturated. Excessive water was drained through entry points of silver wires. The whole was repeatedly shaken to remove any air bubbles that might have appeared. The samples were left to settle and dry in ambient temperature for a week.

Electrical measurements were performed on Biologic SP-300 potentiostat. Cyclic voltammetry was measured in -5V/5V potential window with a scan rate of 100mV/s. Electrochemical impedance spectroscopy (EIS) was measured in the 7MHz – 100mHz frequency window, with 50mV AC perturbation.

To perform signal mixing in the computing concrete system, two separate arbitrary signals from a dual-channel arbitrary waveform generator (TG5012, Aim-TTi, UK) were applied via the WA301 waveform amplifier (Aim-TTi, UK) and impedance matching baluns (1VP-C, Top-View Tek, China) to two chosen terminals of the sample as indicated in Fig. 1. One channel was tuned to 300Hz with sinusoidal wave shape, whereas the second channel was tuned to 290Hz, 280Hz and 275Hz with three different wave shapes for each of these frequencies (sinusoidal, triangular and square). In that scenario, the sinusoidal signal could be perceived as a base probing signal to classify the second signal of unknown shape in a classification task. Both signals were $10V_{pp}$ in amplitude. Processed waveforms were recorded on digital oscilloscope (DSO-X2014A, Agilent Technologies, USA). Examples of recorded time series are shown in Figs. S1-S2 (electronic supporting info).

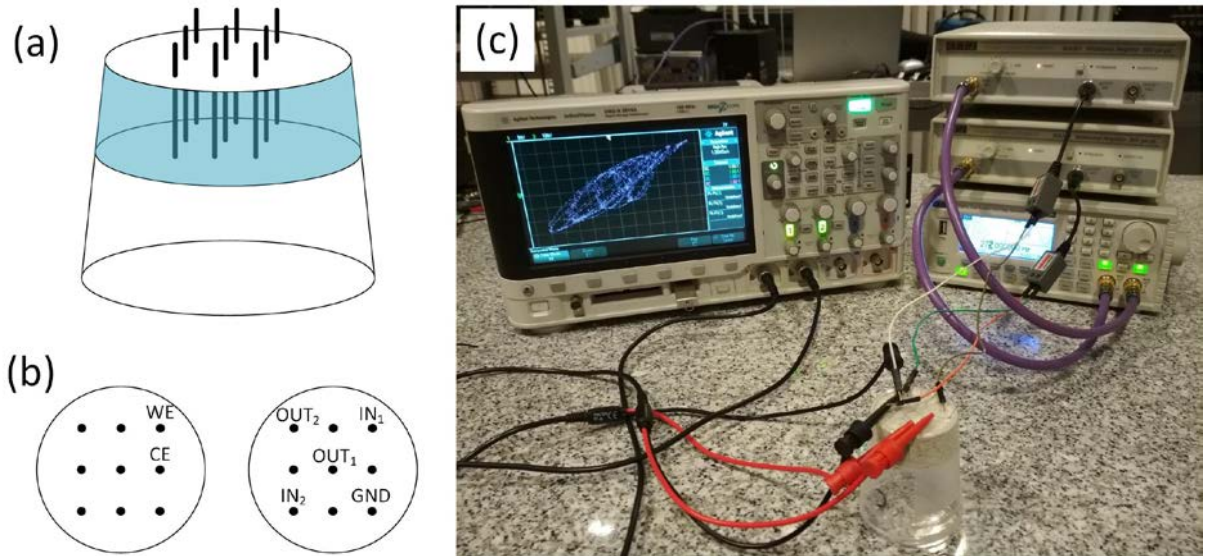

*Figure 1. Schematic view of the computing concrete sample (a), pinouts for voltammetric (left) and signal processing experiments (b) and a real photo of an experimental setup (c). WE and CE stands for working and counter electrodes, respectively. $IN_1$ and $IN_2$ are signal input connectors, $OUT_1$ and $OUT_2$ output ones, GND is a common ground.*

Signals recorded at $OUT_1$ terminal were of higher quality, less scattered and were used for further processing. Only in one case (Petrosian fractal dimension) the $OUT_2$ singlas were used along with $OUT_1$ ones. Signal processing and analysis were performed in Python. Nolitsa module was used for time delay (Delayed Mutual Information method) and embedding dimension (False Nearest Neighbours and Average False Neighbours) estimation. By using Nolds (Python module for nonlinear dynamics study) Correlation Dimension, maximum Lyapunov exponent and Detrended fluctuation analysis (DFA) scores were calculated. Further study of dynamical parameters (Petrosian and Katz fractal dimensions, as well as sample and approximate entropy) was performed using EntroPy Python module for a one-dimensional time series analysis. All analysis was carried out for normalized time series.

3. Results and discussion

Initially, all obtained samples have been characterized with cyclic voltammetry within ±5V window. All samples have shown moderate conductivity and currents up to 2.5 mA have been recorded for samples doped with both semiconducting nanowires and metal shavings (Fig. 2). It was found that undoped concrete as well as concrete with low content of any dopant shows predominant capacitive hysteresis loop (characteristic for ferroelectric materials) [45-48] superimposed on Ohmic current. This bevaviour should be expected for mixed oxide materials [49, 50]. The strongest features charatcteristic for ferroelectric materials has been observed in the case of 10% of SbSI admixture, which is fully consistent with pronounced ferroelectric properties of this material [51-53], but this nonideal capacitive behavior was observed in the majority of cases, the complex character of I/E curves may be interpreted in terms of mixed ferroelectric/antiferorelectric character of studied samples [54]. In light of complex chemical and phase structure of samples this may be fully justified. Detailed analysis of these phenomena is, however, out of scope of this study. In just few cases memristive behavior was observed, with the most pronounced resistive switching in the case of 10% SM sample (Fig 2). Therefore this combination of both dopants was selected for further investigations and for the reservoir computing experiments.

Capacitive properties of selected samples were further addressed using impedance spectroscopy. The junction capacitances are low, which can be seen as a decrease of impedance at high frequency region. This effect is less pronounced for doped materials. Moreover it was found that the Ohmic component increases with increasing concetration of the dopant (Fig. 3a).

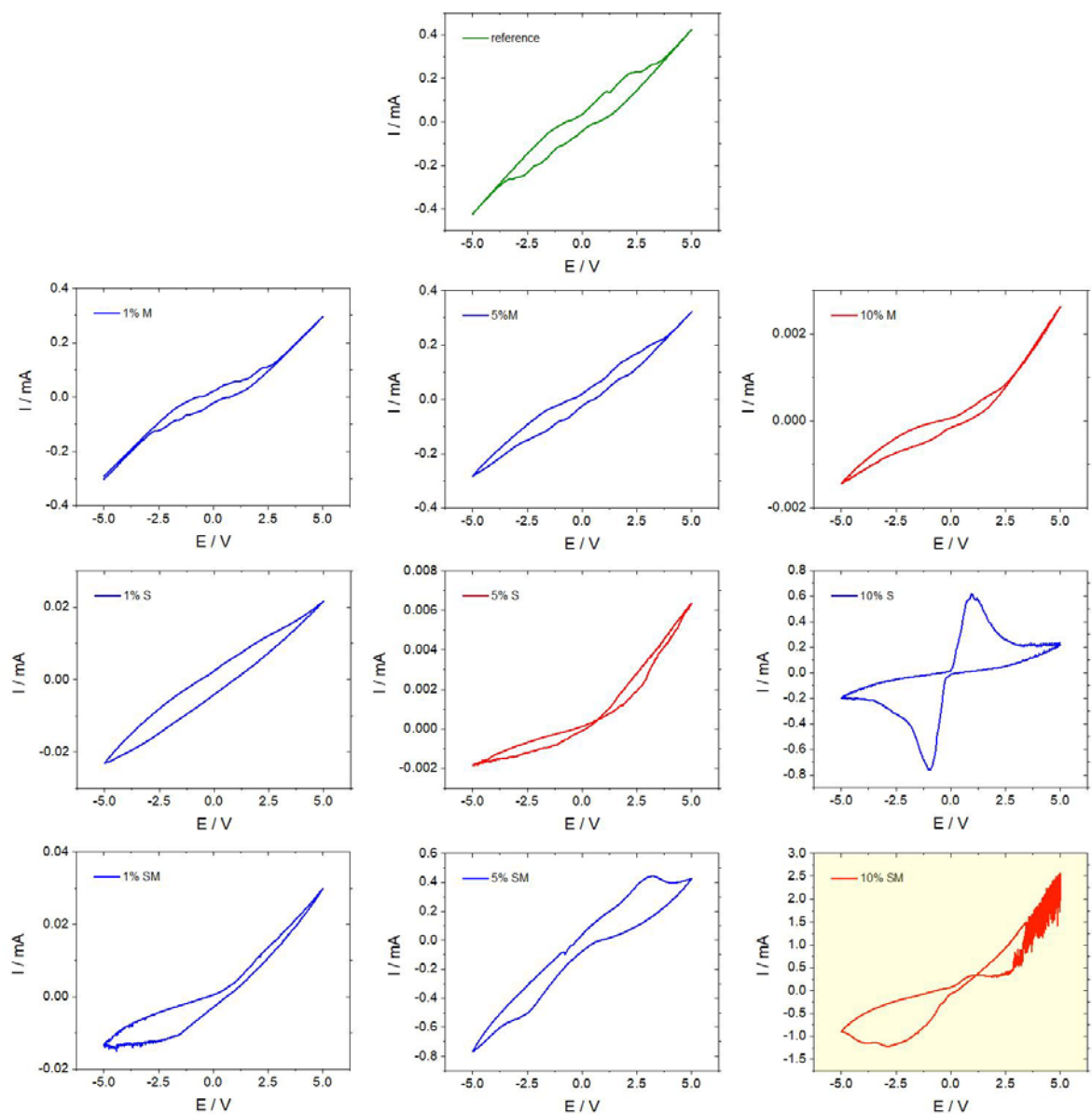

*Figure 2. Voltamperometric characteristics of unoped concrete sample (top, dark green) and concrete containing various amounts of dopants: M – metal shavings, S – antimony sulfoiodide nanowires, SM – 1:1 mixture of both dopants. Some samples show pinched hysteresis loops typical for memristive devices (red) whereas the others are of capacitive character (blue). The most pronounced memristive behavior was observed in the case of 10% SM sample, highlighted in yellow.*

Furthermore, undoped contrete show relatively high phase shift angle at low frequencies, which can be associated with a Warburg impedance related to a slow diffusion process within ceramic matrix. Increasing concetration of dopant reduces this contribution, because other transport mechanisms start to dominate (Fig. 3b).

Based on registered signals, further information processing and analysis was performed based on several methods mentioned *vide supra*. Due to the lack of control over the spatial arrangement of 3d semiconductor/metallic grains suspended in a cement matrix, geometric change of the place from which we read the signals also changes to some extent the calculated parameters. For this reason, the signal readout layer must be properly calibrated to enable signal classification.

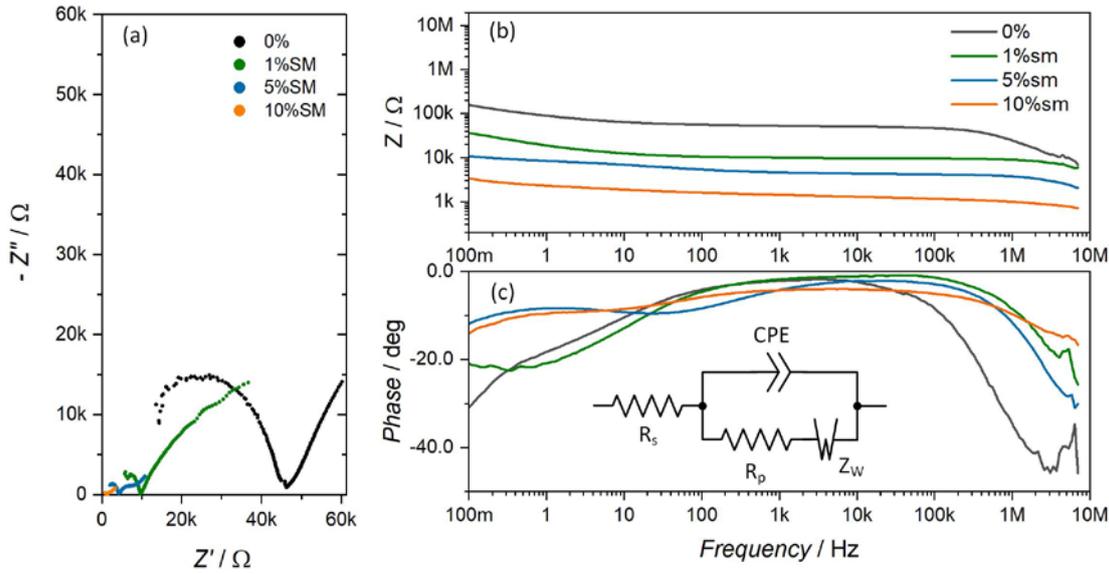

*Figure 3. Impedance spectra of undoped and metal+semiconductor doped concrete samples: Nyquist plot (a) and Bode plots: (b) and phase shift angle (c). A simplified equivalent circuit is also shown. The linear Warburg component at low frequencies is visible only in the case of undoped sample, whereas increased doping is correlated with a decrease of impedance results as well as with significant curvature of the low frequency arm in Nyquist plots.*

*Estimation of time delay and embedding dimension parameters*

At first, Augmented Dickey-Fuller (ADF) test was calculated to check data stationarity. Results show that for a sample size T = 500 the critical values were not exceeded in any case, the highest p-value was obtained for sin/square pair (no more than 1.25%), which means that the null hyphotesis can be rejected (that the data posess "unit root" - presence of stochastic trend) and the time series are in fact stationary [55].

According to the Taken's theorem (which was also shown independently by Packard et. al [56]), single time series can be used to reconstruct so called "delay-coordinate map" based on choosen displacement (time delay) [57-59]. Reconstructed attractors posses the same mathematical properties (e.g. Lyapunov exponenst, fractal dimensions of the attractor or eigenvalues of a fixed point) as the original manifolds of a given dynamical system (usually obtained on the basis of set of ordinary differential equations). It basically comes down to the proper selection of a set of the adjacent coordinates with equal time offset between them. Classical methods of determining the time delay measure the independence of subsequent points in the phase space. Basically for infinite, noise-free time series, the selection of time delay can be choosen almost arbitraty [57], but for experimental data, it is good practice to determine its appropriate value. The time delay for the unfoldment of the attractor was estimated using the Delayed Mutual Information (DMI) [60] and Autocorrelation methods [59].

By applying information theory (for which Shanon provided mathematical formalism [61]) to strange attractors, we can quantify the degree of "surprise" new message provides - in the case of attractors these messages are in the form of values given attractor will take during measurement. The DMI method is based on the quantitative approach to uncertainty about time delayed coordinates given the measure of a chosen coordinate. First minima of calculated functional of joint probability distribution indicates the suitable τ value (Fig 4a). In turn, first zero of autocorelation function gives proper time delay. The autocorrelation methos yields the most suitable delay value of 3, whereas the DMI method, which is more reliable, yields τ = 4. To inspect the validity of the calculated τ value, several delay times were used to reconstruct attractors in the phase space for a randomly selected data set (Fig. S1). It can be observed, that τ = 4 is optimal for the unfoldment of the attractors. It is good practice to choose the smallest time delay required, to avoid phenomena called the irrevalence and redundancy [59, 62]. Irrevalance occurs when the reconstructed attractor fold over on itself thus making it more complicated than original manifold, whereas redundancy means the concentration of attractor shape on the diagonal set. The plot of delayed mutial information versus time delay (Fig. 4a) clearly indicates significant chaotic character of all recorded time series with a contribution of stochastic component. These curves present oscillatory character (fingerprint of chaotic character) and a steep slope as small τ values (stachasticity fingerprint) [63].

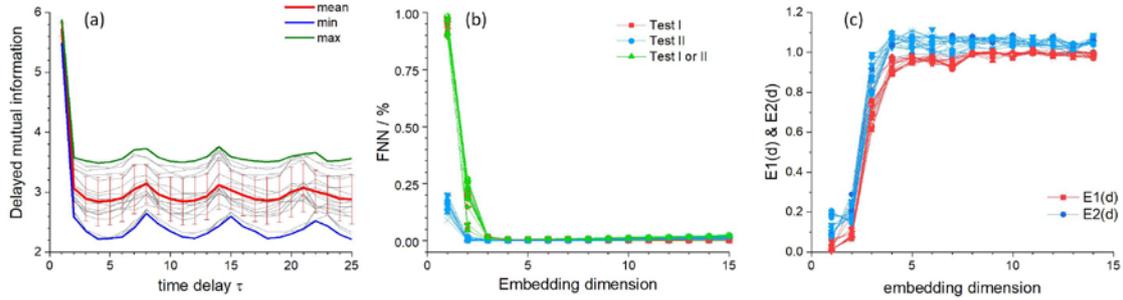

*Figure 4. Results for time delay and embedding dimension calculations for all time series recorded for pristine and doped (10%MS) concrete samples. Graphs present results from Delayed Mutual Information approach (a), False Nearest Neighbours test (b) and Average False Neighbours method (c). Calculated time delay τ = 4 (first minima of DMI, averaged over all data sets), whereas suitable embedding dimension equals four (0% of FNN in all tests and saturation of E1 & E2 in AFN). Descriptions of test criteria can be found in the text.*

Based on a calculated time delay, time-delay embedded trajectories have been plotted (Figs. 5-6) [63]. On both sets of trajectories a highly complex system dynamics can be observed. Frequency ratio of applied stimulation influences irregularities in observed traces, which is represented in beats present in the waveforms (Fig S2, S3) and recurring decimal in these frequency ratios. For 300Hz/290Hz, recurrence of decimal place is observed for 28 digits, for 300Hz/280Hz for seven digits and for 300Hz/275Hz frequency ratio for two digits. The attractors are more regular for the cases where there is a smaller number of periodic digits, as well as for a smaller period of observed beats in the registered waveforms. Moreover, with the progressive deviation from the shape of the basic sinusoidal signal, more and more irregular trajectories can be observed (which may be associated with a greater number of harmonic components of the triangular and square wave signals).

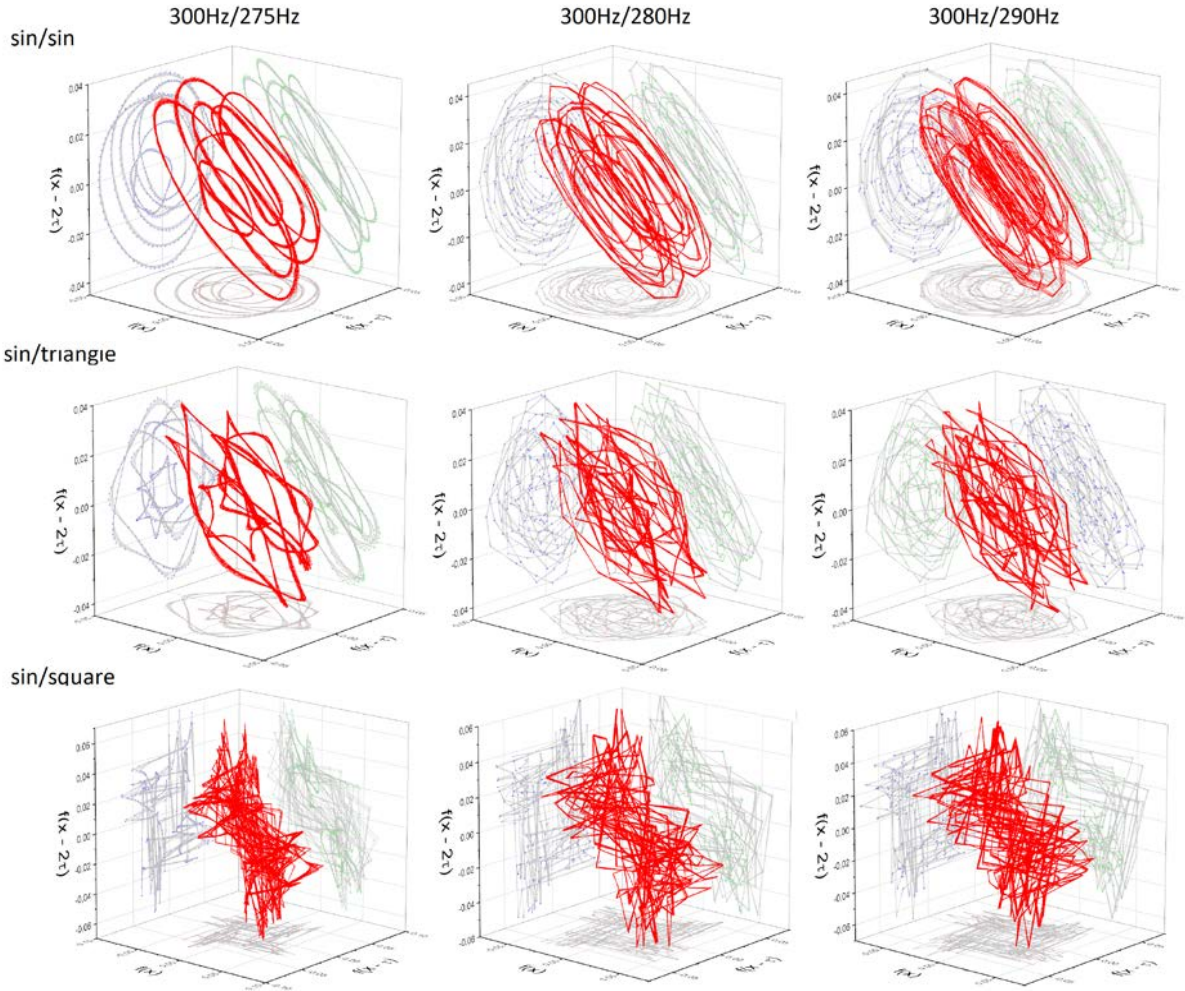

*Figure 5. Embedded time-delay trajectories of time series recorded for un-doped concrete sample for various input waveforms and frequencies, constructed with time delay τ = 4. It should be noted that the f(x) nv f(x-τ) projections are free from diagonal distorsions, which supports the evaluated τ value.*

In order to analyze the nonlinear dynamics of the recorded time series, the appropriate embedding dimension was determined using the False Nearest Neighbour (FNN) method proposed by Kennel et. al. [64] (Fig. 4b) and Average False Neighbours (AFN) method proposed by Cao [65] (Fig. 4c). The FNN method tests whether neighboring points of a specific trajectory in a given embedding dimension are actually neighbors due to the system dynamics or whether they are next to each other only because of the insufficient dimensionality of the phase space. By examining how the number of neighbors changes as a function of dimension, one can determine the appropriate embedding dimension for further analysis. To check the percentage of false neighbors relative to real neighbors, three criterions are used - the first criterion increases the embedding dimension and tests the ratio of Euclidean distance between pairs of points compared to the distance between points with previous embedding dimension value, the second criterion compare relation between reconstructed attractor in higher dimensions and its original size, whereas third criterion uses both previous tests. Both criteria are compared to a heuristicly chosen threshold, values of which are suggested in original work of Kennel et. al. The second condition tries to eliminate the situation where the limited amount of data and the noise present in them causes that the points that are not next to each other are treated as neighbors. To overcome possible problems with choosing proper threshold values in FNN test, Cao proposed his modified FNN method, called AFN or Cao's test. The main difference is that instead of calculating relative distance ratios separately, mean value of all of these distances are analysed between subsequent increasement of embedding dimension (E1(d) in Fig. 4c). Cao further defines another testing criterion (E2(d) in Fig 4c), where ratio of mean distances between subsequent embedding dimensions is calculated for the time delayed one dimensional time series and not for reconstructed vectors as in E1(d) criterion. Previously estimated time delay from DMI and autocorellation methods was used to form time delayed vectors needed in FNN and AFN methods.

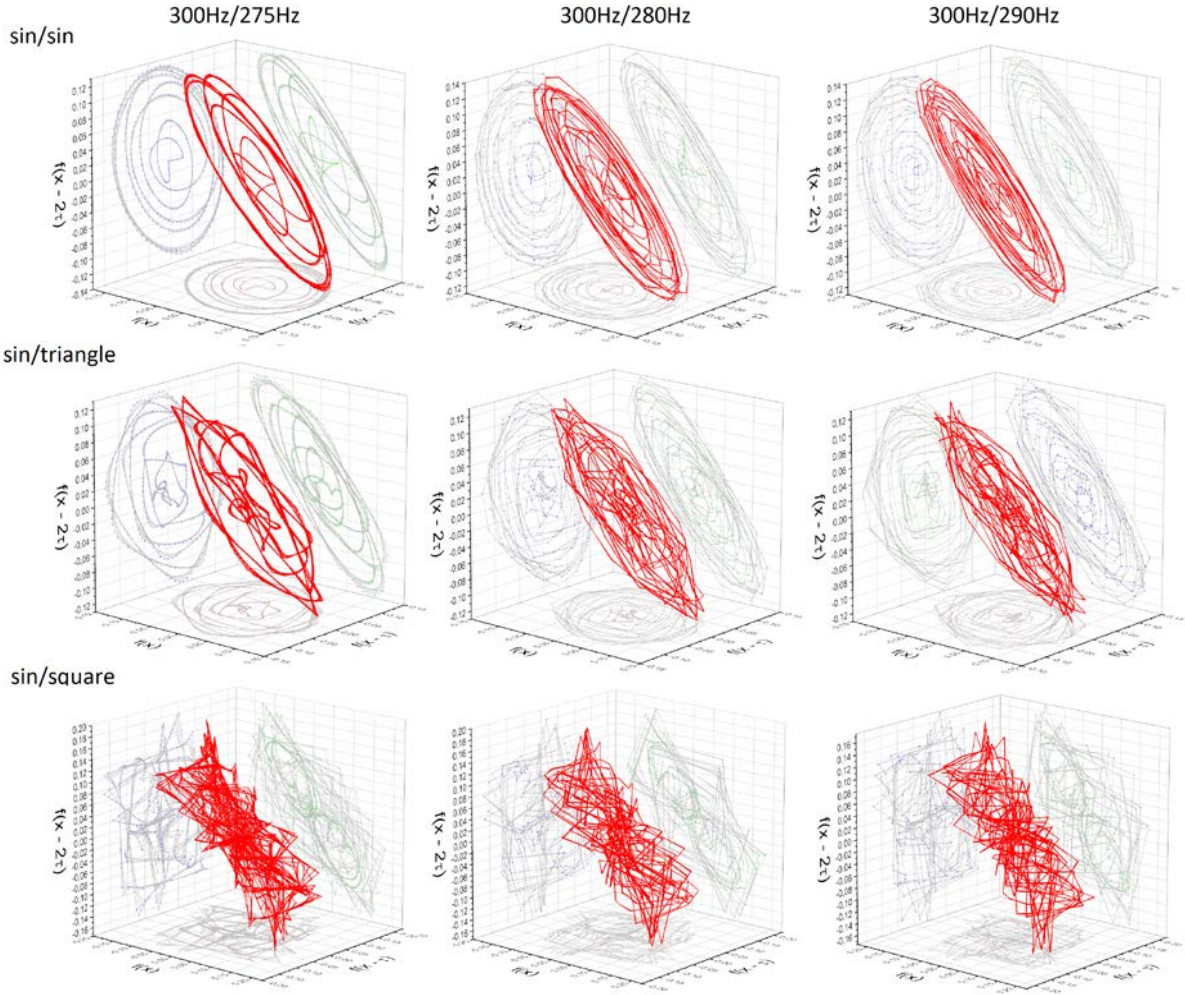

*Figure 6. Embedded time-delay trajectories of time series recorded for doped sample (10% SM) for various input waveforms and frequencies, constructed with time delay τ = 4. It should be noted that the f(x) nv f(x-τ) projections are free from diagonal distorsions, which supports the evaluated τ value.*

FNN results show that the number of false neighbors for all test criteria drops to 0% starting from embedding dimension of 4 (Fig 4b). This outcome is consistent with the results obtained by the AFN method, where both criteria - E1 and E2 - reach saturation starting from the same embedding dimension as in the one indicated in the FNN test (Fig 4c). For this reason, further analysis of nonlinear dynamics was made using the embedding dimension of 4. For a practical reason, however, the attractors are depicted for embedding dimension of 3 (Figs. 5-6). These figures can be considred as 3D projections of 4D attactor obtained by the removal of the 4$^{th}$ coordinate.

The complex character of the recorded time series was further characterized with nonlinear dynamics methods (largest Lyapunov exponent), self-similarity methods (detrended fluctuation analysis, fractal dimensions) and disorder-based methods: dynamic (sample entropy) and structural (permutation entropy) entropy-based methods [66].

*Analysis of non-linear dynamics*

Lyapunov exponents are one of the main indicators of chaos in the study of data possessing non-linear properties [67-69]. It probes the rate of divergence of concomitant trajectories in phase space. The exponential rate of divergence of two chaotic trajectories can be described as follows [70] (1):

$$\Delta(t) \sim \Delta_0 e^{\lambda t}, \tag{1}$$

where $\lambda$ is the Lyapunov exponent, and $\Delta_0$ is the initial separation vector. Due to differences in initial conditions based on a given separation vector, one can obtain a spectrum of Lyapunov exponents. It is common to refer to the largest one as the Maximum Lyapunov exponent (MLE) that is used to probe the predictability and stability of the given data sample. To characterize trajectory instability, MLE can be defined as follows (2):

$$\lambda = \lim_{t \to \infty} \lim_{\Delta_0 \to 0} \frac{1}{t} \ln \frac{\Delta(t)}{\Delta_0} \tag{2}$$

Positive MLE strongly indicates the chaotic nature of system dynamics, especially the sensitivity to the initial conditions, which is known as the "Butterfly effect" [71]. Calculated MLE shows, that for seven cases, un-doped sample presents chaotic behavior (positive MLE) in registered waveforms. In contrast, the doped sample exhibits chaotic behavior in five cases overall (Fig 7). This is generally consistent with the delayed mutual information dependence on time delay (Fig. 4a), which also indicates chaotic features of the recorded time series. Overall increase of MLE can be observed for sine waveforms, whereas overall decrease in MLE is present for the square wave shapes.

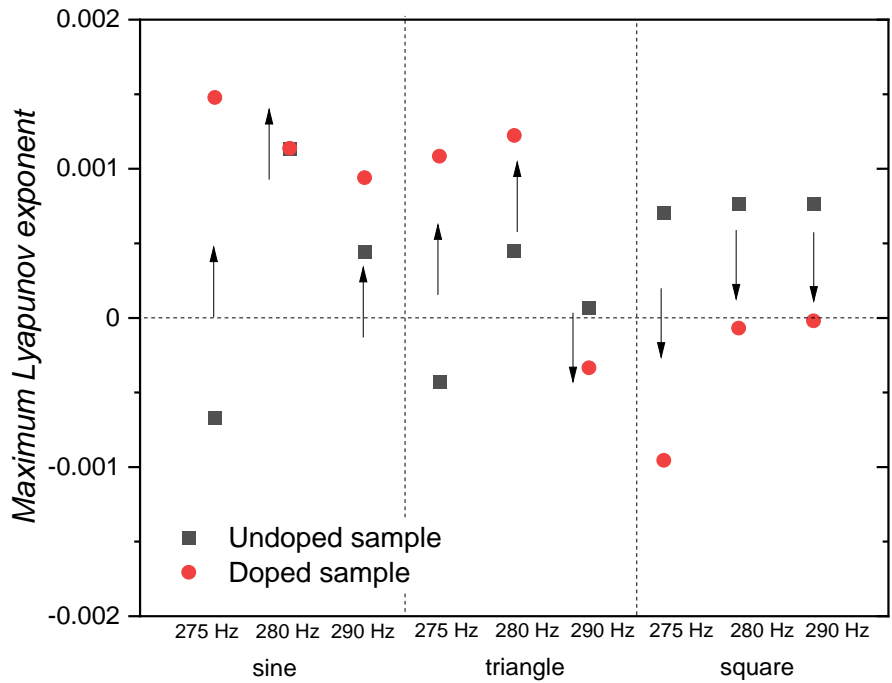

*Figure 7. Maximum Lyapunov exponent for of a time series recorded for different input frequencies/waveforms and for 300 Hz sine drive. Arrows indicate the direction of changes upon doping.*

Detrended Fluctuation Analysis (DFA) is a method for determining the statistical self-affinity of a signal [72, 73]. Self-affinity can be regarded as a property of a fractal time series [74]. Using this parameter, one can easily distinguish whether the stimulated sample was doped or not (Fig. 8). Results indicate correlated ($\alpha > 0.5$) and anti-correlated ($\alpha < 0.5$) character of the registered time series for undoped and doped samples respectively. Both scaling factors $\alpha$ lie between 0 and 1, indicating the stationary character of time series (in accordance with ADF results, *vide supra*). Furthermore, those results indicate the presence of memory in registered time series, [72] which is consistent with the presence of measured memristive traces. Anti-correlated character of time series registered from doped samples may originate from the possibility of flipping resistive state, observed in CV measurements (Fig 2).

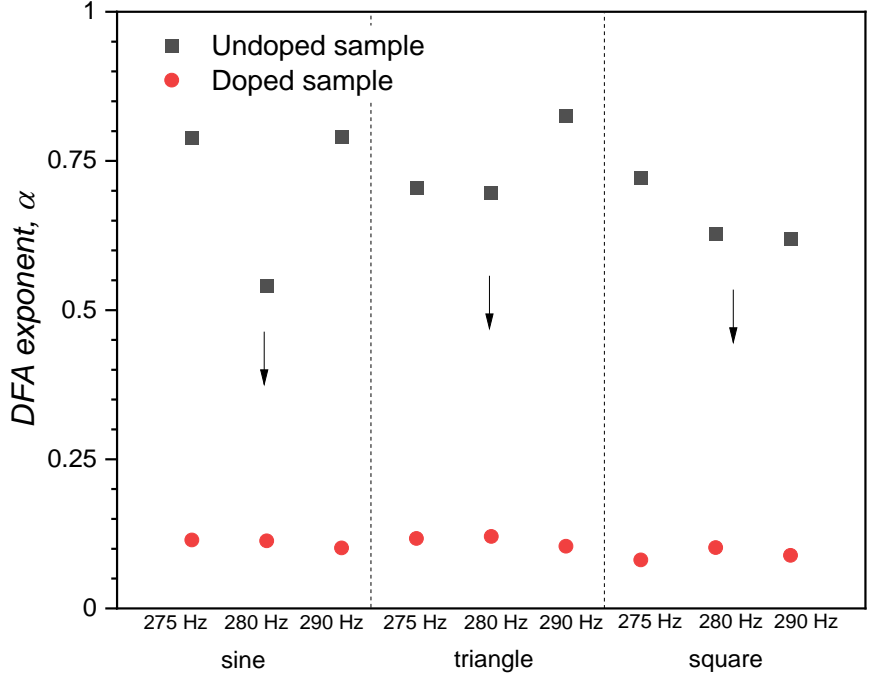

*Figure 8. Detrended fluctuation analysis performed for time series recorded for different input frequencies/waveforms and for 300 Hz sine drive. Arrows indicate the direction of changes upon doping.*

Another parameter used in the study of chaotic and dynamical systems is the correlation dimension ($\nu$) [75]. It is used to probe dimensionality of the space occupied by a set of random points and is often referred to as a type of fractal dimension. For time series of points described as (3):

$$\left\{ \vec{X}(i) \right\}_{i=1} \equiv \left\{ \vec{X}(t+i\tau) \right\}_{i=1}, \tag{3}$$

where $\tau$ is arbitrary, but fixed time increment. The correlation integral is defined as (4):

$$C(r) = \lim_{N \to \infty} \frac{1}{N^2} \sum_{i,j=1}^{N} \Theta\left( r - \left| \vec{X}(i) - \vec{X}(j) \right| \right), \tag{4}$$

where $\Theta(X)$ is a Heaviside step function. For small number $r$, correlation integral behaves according to a power law (5):

$$C(r) \sim r^\nu, \tag{5}$$

where $\nu$ is interpreted as a fractal dimension [75, 76]. As can be seen in Figure 9, the change of correlation dimension is strongly correlated with the shape of mixed signals. The correlation dimensiton is only shifting upwards for the mixing of sin/sin signals, only upwards for triangle signals, whereas for square (280-290 Hz) it shifts downwards and for 275Hz $\nu$ shifts upwards. Based solely on this fact, classification of signal shape can be performed (overall $\nu$ decrease: sine, overall $\nu$ increase: triangle, mixed $\nu$ trend: square).

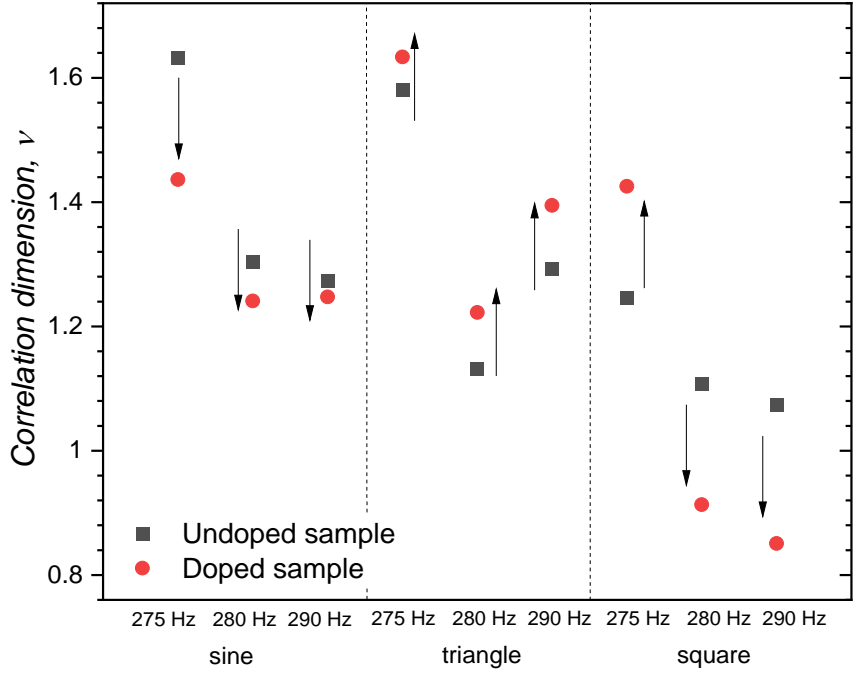

*Figure 9. Correlation dimension of a time series recorded for different input frequencies/waveforms and for 300 Hz sine drive. Arrows indicate the direction of changes upon doping.*

## Classification of waveform on the basis of the decision tree method

As was already mentioned, changing the readout terminal influences obtained dynamical parameters. With this change, analysis of obtained parameters allows for different classification scenario. Analysis of Sample Entropy [77, 78] and fractal dimensions (*vide infra*) gives an alternative approach towards signal classification. Sample entropy is a technique used for probing regularity/complexity (unpredictability of fluctuations) of time-series signals. It possesses desirable characteristics in the form of data length independence and a relatively trouble-free implementation. It is defined as a negative natural logarithm of conditional probability between distances of two sets of points taken from template vector which acts as representation of given data sample. For time series (6):

$$N = \{x_1, x_2, x_3, ..., x_N\} \tag{6}$$

the template vector takes a form of (7):

$$X_m(i) = \{x_i, x_{i+1}, x_{i+2}, ..., x_{i+m-1}\}, \tag{7}$$

where *m* is embedding dimension. Based on this, sample entropy can be described as (8):

$$S_S = -\ln\frac{A}{B}, \tag{8}$$

where *A* and *B* are numbers of template vector pairs having distance ($d[X_{m+1}(i), X_{m+1}(j)]$

and $d[X_m(i), X_m(j)]$, for *A* and *B* respectively) lower than given tolerance *r* (which is taken as a factor of standard deviation). If analyzed data is ordered, then templates for *m* points are also similar for *m*+1 points, and *A*/*B* approaches unity [78]. In that case, negative logarithm will approach 0. Results show, that in most cases (apart from 300Hz/290Hz sin-square), obtained timeseries are more ordered for doped sample, which may be associated with less noise present in the signal (cf. Figs. S1-S2).

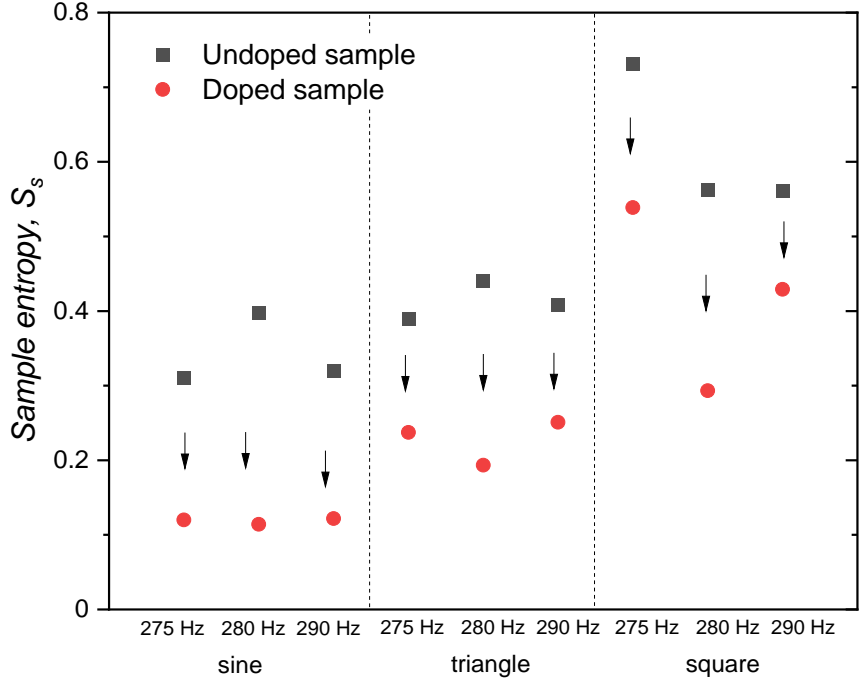

*Figure 10. Sample entropy for different input frequencies/waveforms and for 300 Hz sine drive. Arrows indicate the direction of changes upon doping.*

Trends observed in sample entropy changes (Fig. 10) do not allow unambiguous classification of waveforms, therefore other criteria must be used in parallel.

Permutation entropy is considered as a natural measure of time series complexity via reconstruction of a phase space of any dynamic system [79, 80]. Here it was calculated according to Yan et al. [81] according to the Takens–Maine theorem. The phase space of a time series $\{x(i), i=1,2,3,...,N\}$ can be reconstructed as (9):

$$\begin{cases} X(1) = \{x(1), x(1+\tau), ..., x(1+(m-1)\tau)\} \\ ... \\ X(i) = \{x(i), x(i+\tau), ..., x(i+(m-1)\tau)\} \\ ... \\ X(N-(m-1)\tau) = \{x(N-(m-1)\tau), x(N-(m-2)\tau), ..., x(N)\} \end{cases}, \quad (9)$$

where *m* is the embedded dimension and *τ* is the time delay. Then, the *m* number of real values contained in each $X(i)$ can be arranged in an increasing order as (10):

$$\{x(i+(j_1-1)\tau) \leq x(i+(j_2-1)\tau) \leq ... \leq x(i+(j_m-1)\tau)\}. \quad (10)$$

If there exist two or more elements in $X(i)$ that have the same value, e.g. $x(i+(j_1-1)\tau) = x(i+(j_2-1)\tau)$, their original positions can be sorted in such a way that for $j_1 \leq j_2$ the relation $x(i+(j_1-1)\tau) \leq x(i+(j_2-1)\tau)$ will be obtained. Hence, any vector $X(i)$ can be mapped onto a group of symbols (11):

$$S(l) = (j_1, j_2, ..., j_m), \tag{11}$$

where $l = 1, 2, ..., k \leq m!$. $S(l)$ is one of the $m!$ symbol permutations, which is mapped onto the $m$ number symbols $(j_1, j_2, ..., j_m)$ in $m$-dimensional embedding space. If $P_1, P_2, ..., P_k$ are used to denote the probability distribution of each symbol sequences, respectively, and the condition (12):

$$\sum_{l=1}^{k} P_l = 1 \tag{12}$$

is fulfilled, the permutation entropy of a time series $\{x(i), i = 1, 2, 3, ..., N\}$ can be defined as a Shannon entropy for the $k$ symbol sequence (13):

$$S_p(m) = -\sum_{l}^{k} P_l \ln P_l. \tag{13}$$

As the maximum value of $S_p(m)$ for a uniform probability distribution is equal to $\ln m!$, it is usually given as a normalized value (14):

$$S_p(m) = \frac{-\sum_{l}^{k} P_l \ln P_i}{\ln m!} \tag{14}$$

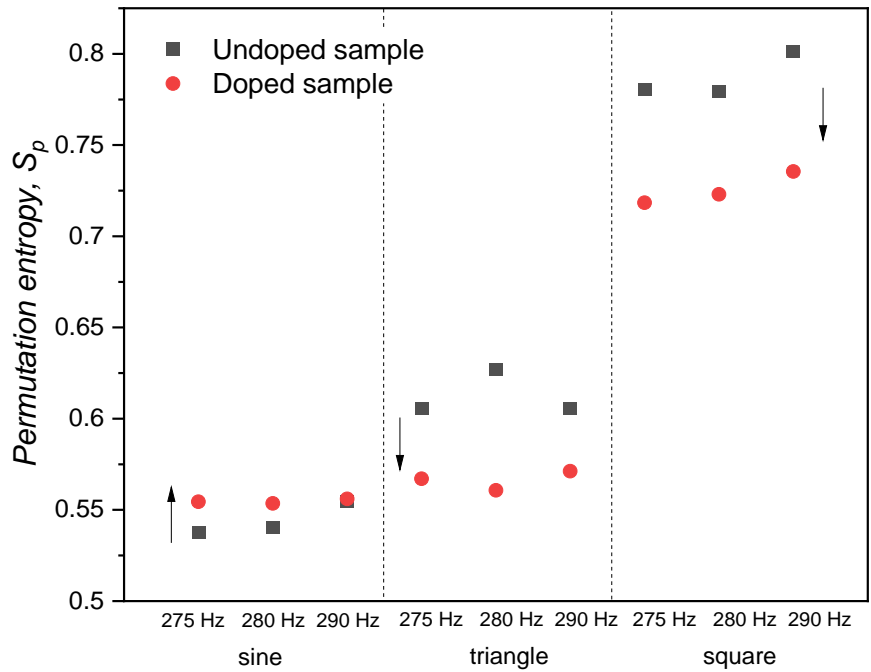

*Figure 11. Permutation entropy for different input frequencies/waveforms and for 300 Hz sine drive. Arrows indicate the direction of changes upon doping.*

The values of permutation entropy serve as a measure of time series randomness. Smaller values indicate less chaotic behavior wherear values approaching the unity indicate highly chaotic behavior and thus unpredictability of time series. These data suggest that time series recorded for sine and triangle waveforms are significantly more ordered (i.e. less chaotic) than those for square waves, which may be utilized as a classification tool.

Analysis of Petrosian [82] and Katz fractal dimensions [83] allows a different approach for signal classification. Katz fractal dimension ($D_K$) calculates the fractal dimension of data directly from the waveforms without the need for their abstract representation. It is defined as (15):

$$D_K = \frac{\log_{10} n}{\log_{10} \frac{d}{L} + \log_{10} n},  \quad (15)$$

where $d$ is calculated fractal dimension, $n = L/a$ ($n$ is used for normalization of distances – $L$ is the total sum of lengths of the successive points, and $a$ is averaged distance between successive points) and $d$ is the maximum distance between the first point and any other point within the data set. $D_K$ is known to overestimate probed fractal dimension, hence large differences in obtained $D_K$ and correlation dimension scores [84]. Figure 12 shows calculated Katz fractal dimensions of attractors for various waveform and frequency combinations.

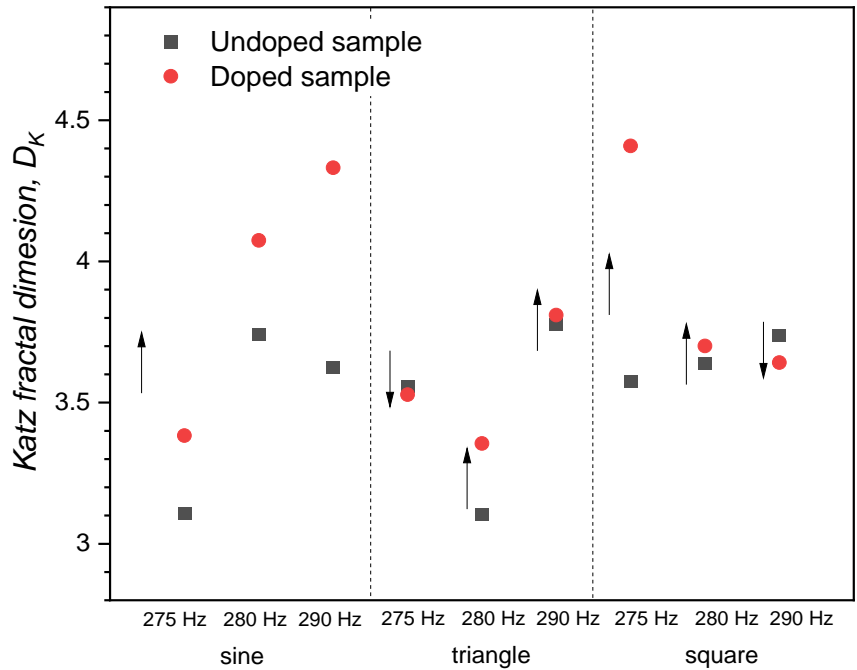

*Figure 12. Values of Katz fractal dimension score for doped and un-doped sample different input frequencies/waveforms and for 300 Hz sine drive. Arrows indicate the direction of changes upon doping.*

Another approach in probing fractal dimension of time series was suggested by Petrosian [82]. Petrosian fractal dimension ($D_P$) is calculated for binarised time series. It is defined as follows (16):

$$D_P = \frac{\log_{10} N}{\log_{10} N + \log_{10} \frac{N}{N + 0.4 N_\delta}}, \tag{16}$$

where $N$ is the length of the time series, and $N_\delta$ is the number of sign changes in the signal derivative. It can be observed in Fig. 12, that Petrosian fractal dimensions increase in the series sine<triangle<square for both output signals and both materials. There is, however a significant change in the undoped/doped difference, as indicated by black arrows in Fig. 12.

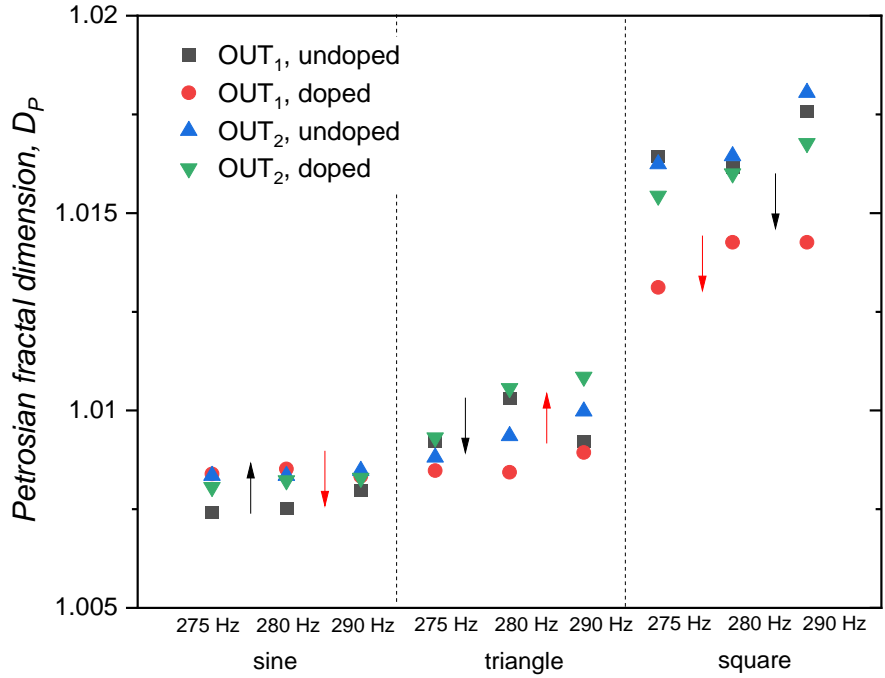

*Figure 13. Petrosian fractal dimension for timeseries collected from $OUT_1$ (a) and $OUT_2$ (b) device terminals. Black arrows indicate trends for $OUT_1$, wherear red arrows for $OUT_2$ (cf. Fig. 1 for terminal markings).*

Observed trends can constitute a set of criteria that can be used for waveform discrimination on the basis of the signal dynamics in pristine and heavily doped concrete blocks. These criteria, along with the dynamic analysis presented above, can be regarded as a readout layed of the reservoir computing system.

*Table 1. A list of trends observed between different shapes of mixed signals for different methods of analysis. Trends are shown for the doped sample in relation to the un-doped one.*

|  | sine | triangle | square |
| --- | --- | --- | --- |
| Permutation entropy | increases | Decreases | increases |
| Katz fractal dimension | Increases | Mixed | Increases |
| Petrosian fractal dimension | Increases | Decreases | Decreases |

Based on different trends of change of the given parameter between doped and undoped samples, one can classify signal shapes in a decision tree manner. A decision tree could be constructed as follows:
1. If calculated permutation entropy decreases and Katz fractal dimension is of mixed trends, then the signal is of the triangle wave shape.
2. If calculated Petrosian fractal dimension is increasing, then the signal is of sinusoidal shape, if its decreasing (and was increasing in the previous step), then it is of square shape.

As is was already mentioned, changing the readout terminal influences obtained dynamical parameters (*vide supra*). With this change, the analysis of obtained parameters allows for different classification scenarios. Based solely on the Petrosian fractal dimension of registered time series but analysed from two different device terminals (OUT1 and OUT2, Fig 16) another classification variant of a decision tree manner can be obtained.

*Table 2. A list of trends observed between different shapes of mixed signals for different methods of analysis. Trends are shown for the doped sample in relation to the un-doped one.*

|  | sine | triangle | square |
|---|---|---|---|
| Petrosian fractal dimension (OUT1) | Increased | Decreases | Decreases |
| Petrosian fractal dimension (OUT2) | Decreases | Increases | Decreases |

A decision tree based on trends summarized in Table 2 could be constructed as follows:
1. If calculated Petrosian fractal dimension (OUT1) is increasing, then the signal is of sinusoidal shape, if its decreasing (and was increasing in the previous step), then it is of square or tranglular shape.
2. If calculated Petrosian fractal dimension (OUT2) is increasing, then the signal is of triangular shape, if its decreasing (and was decreasing in the previous step), then it is of square shape.

Along with various trends (changes in various dynamic parameters upon transition from pristine to doped concrete) another classification system, based of the whole collection of time series can be also derived (Fig. 14). Three selected criteria provide the best classification of waveforms and also provide means for classification of concrete material. Interestingly, detrended fluctuation analysis yields exponent α which can differentiate between doped and undoped concrete, but does not provide means for signal classification. Time series recorded for pristine concrete are much higher ($\alpha > 0.50$) than for doped concrete ($\alpha < 0.25$). This indicates statistically higher correlation of time series for pristine material and anticorreltion for doped one. This may be associated with quite different dielectric responses of both materials. Sample entropy ($S_s$) is not a useful classification criterion, both due to the same trend over all samples (vide supra) and due to very scattered values (Fig. 13). Petrosian fractal dimension for sine and triangular waveforms are significantly lower than for square signals, therefore it may serve as a crude criterion for detection of square wave signals. Finally, the permutation entropy provides a weak classification tool for all waveform shapes: sine waves yield the lowest values, triangular waves the intermediate ones, whereal square waves the highest values of $S_p$. This criterion should be considered as a fuzzy one, as the boundary between sine and triangular waves is not well defined.

Despite the requirement of a complex numerical processing of the data in order to extract the classification parameters the results presented here clearly indicate, that computation with appropriately prepared concrete blocks is possible. Surprisingly, concrete – one of the most ubiquitous construction materials – shows complex chaotic dynamics when stimulated with acoustic frequency electrical signals. Moreover, this dynamics can serve as a classification tool. Selection of wider range of frequencies and waveforms should lead to more complex classification patterns. It seems, that concrete itself presents internal electrical dynamics so complex, that in principle it shoul be capable of much more complex computational tasks in real time. Recently reported speech recognition in coupled nano vortex oscillators [85] is based on a systes of comparable dynamics (however shifter to radio frequency range). Therefore any flassification of acoustic signals required their mapping into radio frequencies. The system presented here performs complex classification tasks directly on amplified signals

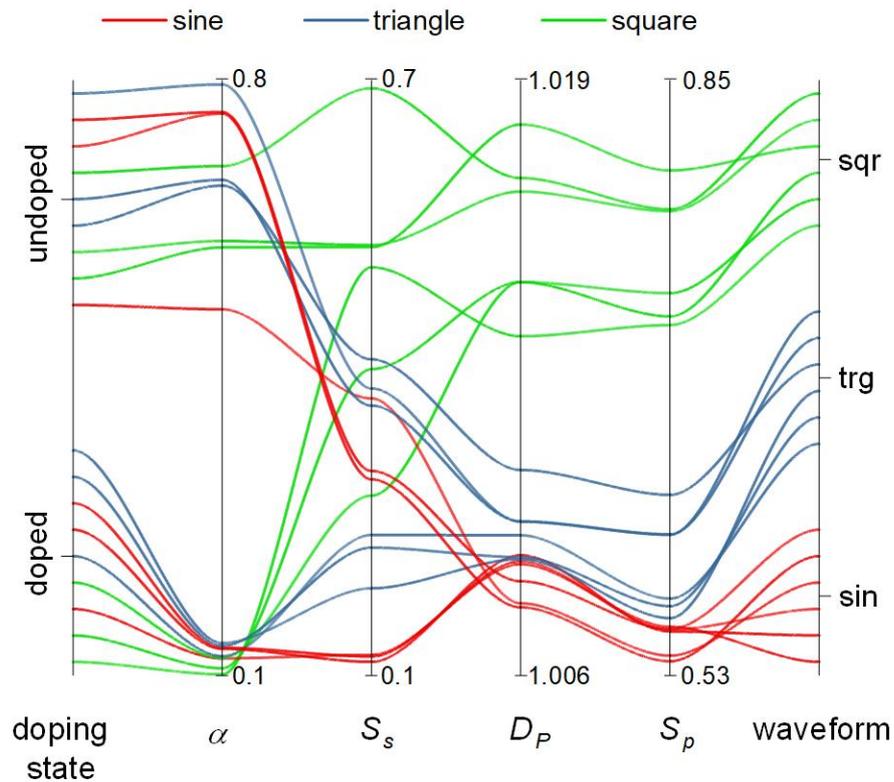

*Figure 14. Parallel coordinate plot for all time series and selected dynamic criteria: detrended fluctuation exponent ($\alpha$), sample entropy ($S_s$), Petrosian fractal dimension ($D_P$) and permutation entropy ($S_p$). Detrended fluxtuation exponent can serve as a classification factor for the concrete doping state, whereas permutation entropy significantly classifies time series according to their waveforms.*

The device presented in this paper (Fig. 15) can be regarded as a heterodic reservoir computing system. The hererodicity originates from combination of in materio reservoir processing of input signals followed by software algorhitms for postprocessing. In a far-fetched vision, an alternative, *in materio*-based readout should be considered, but the complexity of required signal processing seems to exceed the state of the art of *in materio* reservoir computers. The observed features indicate, that the small concrete blocks with silver wire electrodes show a set of features sufficient for reservoir computing. The fading memory feature is represented by capacitive and memristive character of the device, whereas internal dynamics is provided by the drive signal. It shows the echo state property, as the output at selected point reflects features on inner electrical dynamics. The dynamic response of the system is complex enough to provide sufficient separability (in the sense of Stone-Weierstrass theorem) of the input data [86]. It also presents some generalization features, as the observed output space (trends of several criteria) is much smaller than the (infinite) input space of various signals. Due to the specific task and material properties of pristine and doped concrete the output layer, especially the postprocessing part, is relatively complex. It should be noted, however, that this was the requirement for a relatively hard task for memristive reservoir computing systems and that the memristive properties of deliberately chosed materials was very poor. Despite this, the classification task is successfully performed. Future ugrade of this system may involve fuzzy logic inference engine (or multinary logic), as the output trends are not crisp values, and therefore the fuzzy descriptors may be more adequate. Interestingly, multinary and fuzzy logic may be also implemented in related materials [87-89].

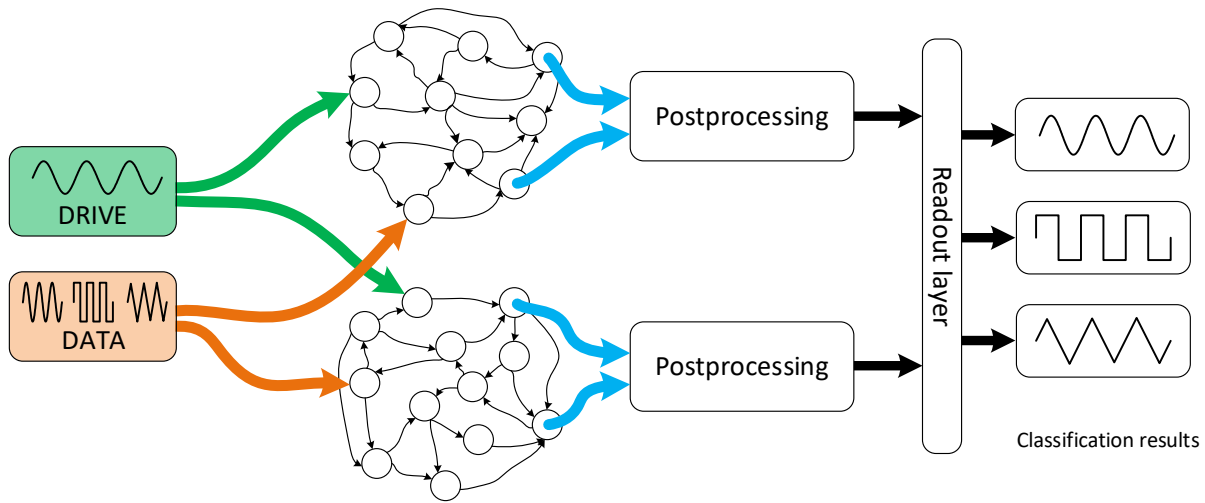

*Figure 15. A scheme of a dual concrete-based reservoir computing system used for waveform classification.*

## 4. Concluding remarks

In this article, the classification of signal shapes was shown based on *in materio* computing concrete hardware system. Samples present a highly non-linear response in regard to data transformation, possess rich configuration state space, and their dynamics (when stimunated with a simple sine wave drive) is represented in the form of four-dimensional attactors. These features make them a suitable platform for reservoir computing implementation. Depending on used terminals for the readout layer, different classification scenarios can be achieved. The presented results can be treated as proof of the concept for the possibility of information processing and classification tasks performed by appropriately doped ubiquitous construction materials. Further development of the concept can bring the realization of more aspects of a multisensory infrastructure capable of information processing based on its embedded hardware and intelling computing houses as a far-fetched vision.

## 5. Acknowledgemens

The authors acknowledge the financial support from the Polish National Science Centre within the MAESTRO and PRELUDIUM projects (grant agreement No. UMO-2015/18/A/ST4/00058 and UMO-2018/31/N/ST5/03112). DP has been partly supported by the EU Project POWR.03.02.00-00-I004/16.

**Electronic supplementary information**

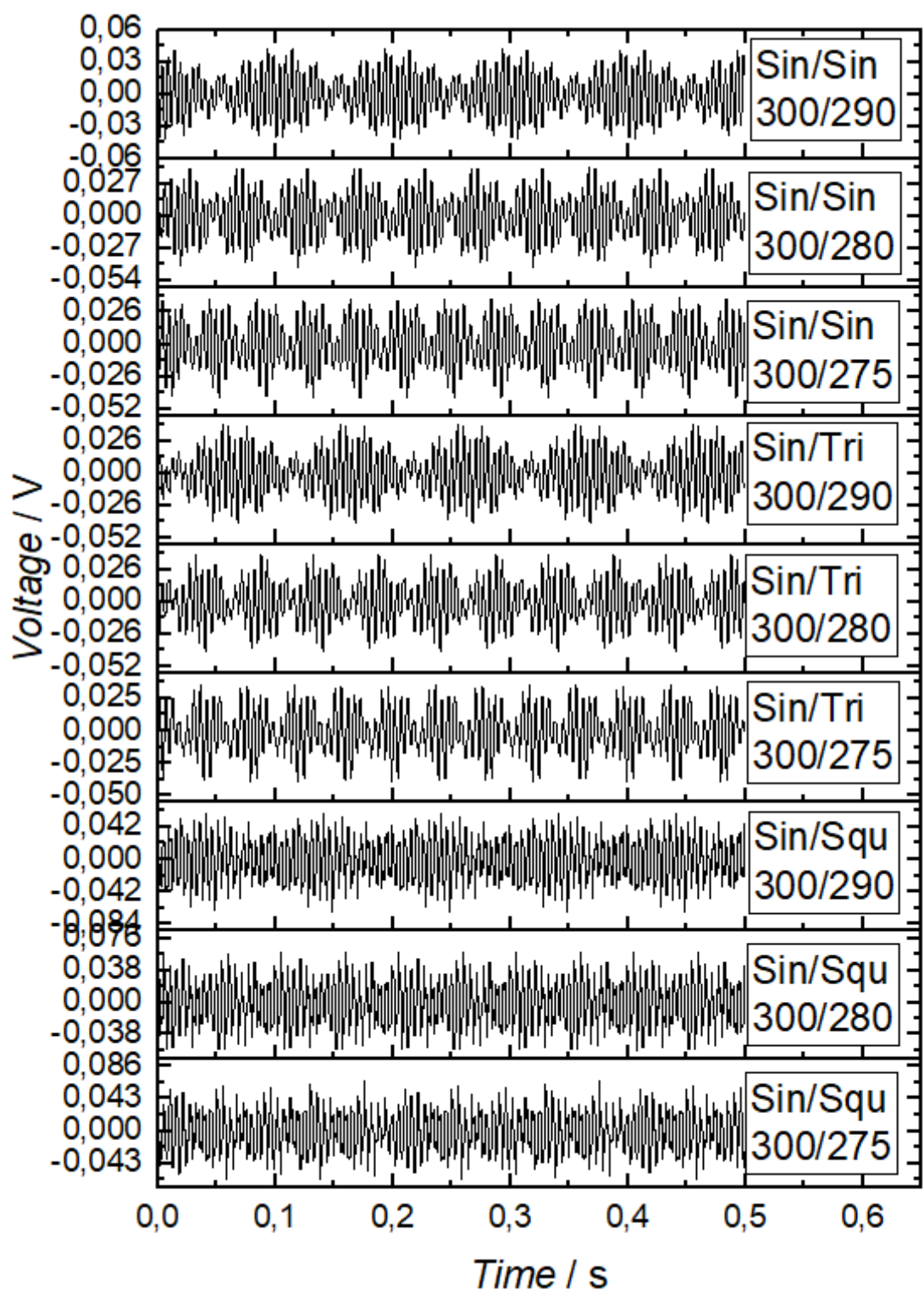

*Figure S1. Registered timeseries for undoped sample (Sin – sinusoidal, tri – triangular, squ – square wave form). Frequencies are given in Hz.*

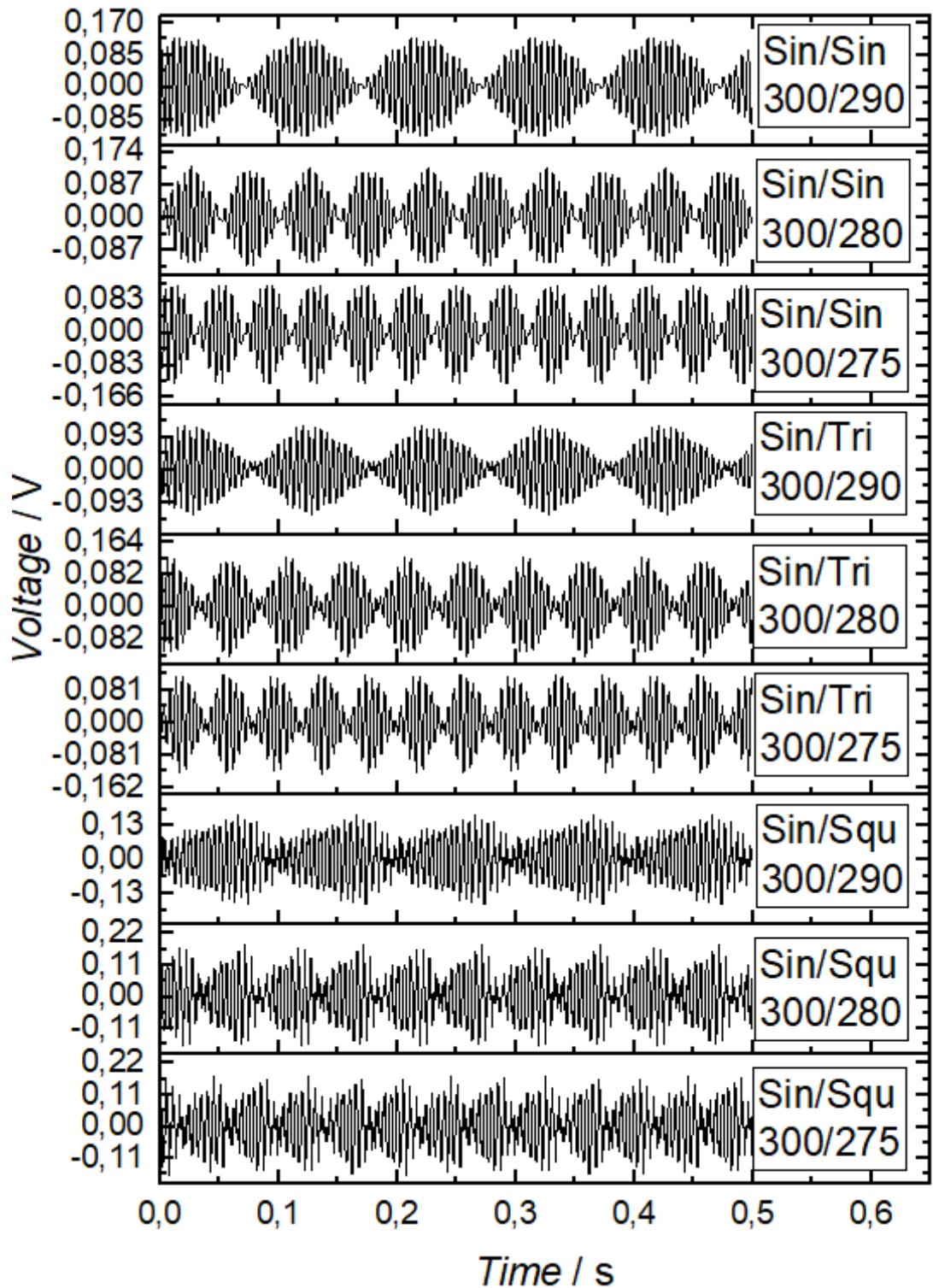

*Figure S2. Registered timeseries for doped sample (Sin – sinusoidal, tri – triangular, squ – square wave form). Frequencies are given in Hz.*

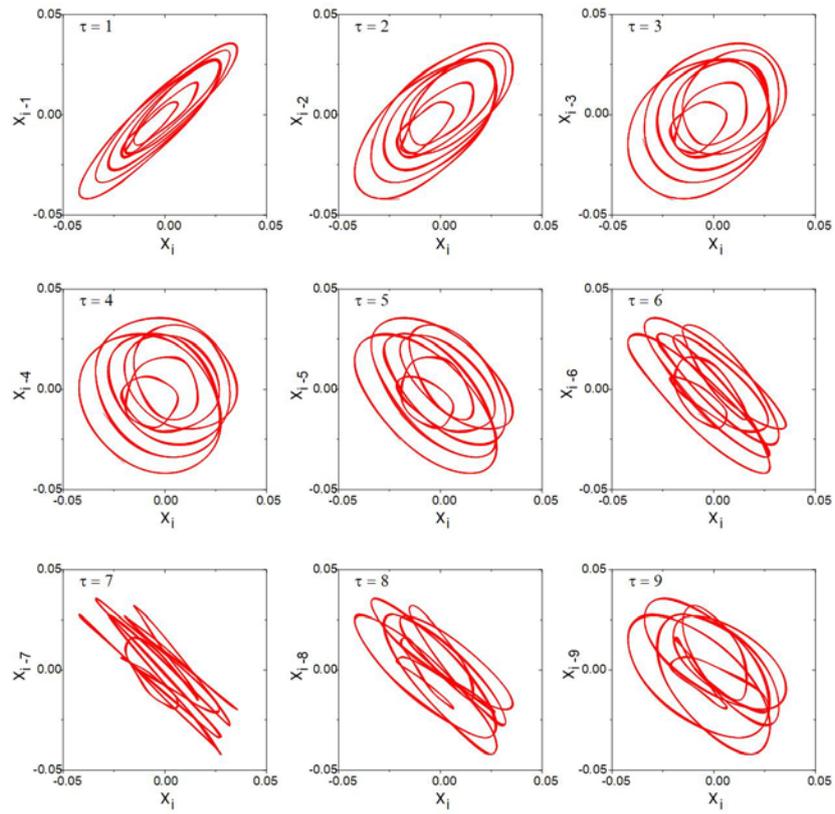

*Figure S3. 2D projections of trajectories (return plots) of the signal obtained for two sine waves (275 and 300 Hz) in undoped concrete. The case of τ = 4 does not show any diagonal stretching.*